\documentclass[aps, prd, twocolumn, superscriptaddress, nofootinbib]{revtex4}
\usepackage{graphicx}
\usepackage{dcolumn}
\usepackage{amssymb}
\usepackage{amsmath,amssymb,amsfonts,bbold}

\newcommand{\plk}{\mathfrak h}
\newcommand{\be}{\begin{equation}}
\newcommand{\ee}{\end{equation}}
\newcommand{\bea}{\begin{eqnarray}}
\newcommand{\eea}{\end{eqnarray}}

\begin{document}

\title{Do we live in an eigenstate of the ``fundamental constants''
operators?}
\author{John D. Barrow}
\affiliation{DAMTP, Centre for Mathematical Sciences, University of Cambridge,Wilberforce
Rd., Cambridge CB3 0WA, United Kingdom}
\author{Jo\~{a}o Magueijo}
\affiliation{Theoretical Physics Group, The Blackett Laboratory, Imperial College, Prince
Consort Rd., London, SW7 2BZ, United Kingdom}
\date{\today }

\begin{abstract}
We propose that the constants of Nature we observe (which appear as
parameters in the classical action) are quantum observables in a kinematical
Hilbert space. When all of these observables commute, our proposal differs
little from the treatment given to classical parameters in quantum
information theory, at least if we were to inhabit a constants' eigenstate.
Non-commutativity introduces novelties, due to its associated uncertainty
and complementarity principles, and it may even preclude hamiltonian
evolution. The system typically evolves as a quantum superposition of
hamiltonian evolutions resulting from a diagonalization process, and these
are usually quite distinct from the original one (defined in terms of the
non-commuting constants). We present several examples targeting $G$, $c$ and 
$\Lambda $, and the dynamics of homogeneous and isotropic Universes. If we
base our construction on the Heisenberg algebra and the quantum harmonic
oscillator, the alternative dynamics tends to silence matter (effectively
setting $G$ to zero), and make curvature and the cosmological constant act
as if their signs are reversed. Thus, the early Universe expands as a
quantum superposition of different Milne or de Sitter expansions for all
material equations of state, even though matter nominally dominates the
density, $\rho $, because of the negligible influence of $G\rho $ on the
dynamics. A superposition of Einstein static universes can also be obtained.
We also investigate the results of basing our construction on the algebra of 
$SU(2)$, into which we insert information about the sign of a constant of
Nature, or whether its action is switched on or off. In this case we find
examples displaying quantum superpositions of bounces at the initial state
for the Universe.
\end{abstract}

\maketitle



\section{Introduction}

It is sombering to recognise that we have never successfully predicted the
values of any fundamental dimensionless constants of Nature, yet we measure
them with our most accurate experiments: they combine our greatest
experimental knowledge with our greatest theoretical ignorance.
Historically, there have been different attitudes and expectations regarding
the numerical values of the constants of Nature. Einstein \cite{ein}
believed there was a single logically self-consistent, rigid, theory of
everything (a \textquotedblleft unified field theory\textquotedblright )
whose defining constants were uniquely and completely specified. The
direction of modern particle physics and cosmology now prefers the opposite
view that there are many self-consistent theories of everything, with
different suites of defining constants of Nature, and we inhabit one of them
as a result of random symmetry-breakings at ultra-high energies. The
question of how much of our known low-energy physics must necessarily reside
in the vacuum state of one of these many possible theories of everything
remains open \cite{swamp}.

These scenarios open up the possibility that the low-energy constants of
physics could be different -- and may even be different in widely separated
places in the universe on both sub and super-horizon scales. An added
complexion arises from the prospect that the quantities we call constants of
Nature may not be the fundamental unchanging constants defining the most
basic theory of everything. Our observed ``constants'' may therefore be
allowed to be time and/or space dependent variables \cite{jdb2} without
undermining the invariant status of the true constants in the most basic
theory. A simple example is provided by 3-d constants in a 3-d submanifold
of a higher-dimensional theory in which any time change in the mean size of
additional dimensions will be seen in the time change of the 3-d ``constants'' 
\cite{3d}, like the fine structure constant and its effects on quasar
spectra at high redshifts \cite{webb}.

We note that all previous attempts to variabilize the constants of Nature
have been classical: one produces an action principle for which one or more
constants (e.g. the particles masses, the coupling constants, and $G$, $c$ or
even $\hbar $) are promoted to dynamical (scalar) fields \cite{BD, B, SBM}.
Any quantization of these fields is banned. In this paper we speculate
further that the set up of \textquotedblleft varying
constants\textquotedblright\ may be purely quantum and devoid of a classical
counterpart. In such a situation the constants that we observe would have
fixed values but they would not need to be unique. That is the speculative
avenue to be explored in this paper.

Usually we start from a classical system (or theory), and then
\textquotedblleft quantize\textquotedblright\ via a given prescription, such
as canonical quantization or the path integral formalism. The possible
pitfalls are well-known. The quantum theory contains more information than
the classical theory, so ambiguities arise, such as ordering issues, inner
product uncertainties, or a multitude of options for the quantum
implementation of constraints, should they exist. Quite often a given
classical system or theory leads to a variety of quantum analogues, no
matter how careful the quantization procedure is, and we know of no way to
generate quantum solutions of a theory directly from solutions of its
classical counterpart.

Another example of the conceptual difficulties plaguing the interaction
between the classical and quantum worlds is found in the opposite direction,
when one seeks a classical or semi-classical limit of quantum theories.
Quantum gravity has been a major source of trouble in this respect, with
mathematically consistent non-perturbative constructions risking being
nothing but a figment of our imagination for want of a suitable (semi-)
classical limit. Even in more mundane situations (and ignoring the notorious
measurement problem) one can say that the quantum and classical worlds sit
together rather uneasily. But could it be that in some situations, such as
near the start of our universe, this interaction happens in a totally novel
way? In this paper we propose that the constants of Nature are quantum
observables in a purely kinematic Hilbert space, and that we are currently
living in an eigenstate of their corresponding operators. An eigenstate
corresponds to a classical lagrangian with fixed constants and a dynamics
which may then be quantized by traditional methods. We present the formalism
in Section~\ref{simplestmod}, allowing for superpositions of such dynamics.

The non-trivial aspects of this proposal start when we elaborate this basic
picture: for example, by allowing two constants to be non-commuting
observables (Section~\ref{noncmod}). Then, a fundamental indeterminacy would
even be built into the classical dynamics. In fact complementarity may
preclude hamiltonian evolution altogether. We illustrate these points with a
toy model in Section~\ref{ho}: a simple harmonic oscillator for which the
mass and the spring constant are non-commuting \textit{constant}
observables. Depending on technical assumption, a diagonalization procedure
may be possible, leading to alternative dynamics. The general evolution of
the system is then a quantum superposition of these qualitatively different
evolutions.

For the rest of the paper we transpose this construction to cosmology,
targeting $G$, $c$ and $\Lambda $, and the dynamics of homogeneous and
isotropic Universes. In Sections~\ref{hocosmo},~\ref{generalgamma} and~\ref%
{Lambda-mods} we examine what happens if $G$ and $c$ are complementary
variables. We find that, even though matter dominates the spatial curvature $%
K$ and $\Lambda $ early on, it does not affect the expansion of the
Universe, which can be a general superposition of Milne (if $K=1$ and $%
\Lambda =0$) or de Sitter (if $K=0$ and $\Lambda <0$) expansion profiles. In
general the universe evolves as a quantum superposition of dynamics ruled by
effectively setting $G$ to zero, quantizing the speed of light, and
reversing the effective signs of $K$ and $\Lambda $. A similar pattern can
be obtained by making $c$ and $\Lambda $ a complementary pair (Section~\ref%
{candL}). Superpositions of stable static universes also appear as solutions
(Section~\ref{static}).

Instead, in Section~\ref{su2} and~\ref{su2cosmo} we base our construction on
the algebra of $SU(2)$, into which we insert information about the sign of $%
G $, $\Lambda$ and $K$, or whether their action is switched on or off. We
find examples displaying quantum superpositions of bounces at the initial
state for the Universe.

Finally, in two concluding Sections we discuss the physical meaning of our
construction, summarise its main results and outline some of its open
problems,

\section{The simplest model}

\label{simplestmod}

Let us consider a classical dynamics defined by an action, $S$, which is a
functional over generic variables (or fields) globally denoted by $x$, and
that depends on $n$ \textquotedblleft constants\textquotedblright ,\ $\kappa
_{i}$: 
\begin{equation}
S=\int \mathcal{L}(x;\kappa _{i}).
\end{equation}%
The associated hamiltonian (obtained while keeping $\kappa _{i}$ constant)
is: 
\begin{equation}
\mathcal{H}=\mathcal{H}(x,p;\kappa _{i})=\mathcal{L}-pq,
\end{equation}%
with $x$ and conjugate momenta $p$ satisfying a diagonal Poisson bracket: 
\begin{equation}
\{x,p\}=\delta ,
\end{equation}%
(where the Dirac/Kronecker $\delta $ involves all indices and variables on
which the $x$ and $p$ depend). First, we construct a Hilbert space $\mathbb{H%
}$ in which the $\kappa _{i}$ are represented by Hermitian operators, $\hat{%
\kappa}_{i}$ (\textquotedblleft observables\textquotedblright ), with
eigenstates associated with eigenvalues $\kappa _{i}$ for the constants: 
\begin{equation}
\hat{\kappa}_{i}|\kappa _{i}\rangle =\kappa _{i}|\kappa _{i}\rangle .
\end{equation}%
Since this is a theory of these \textquotedblleft
constants\textquotedblright , no dynamics (or quantum hamiltonian) is given
to Hilbert space $\mathbb{H}$, i.e. it is a purely kinematical Hilbert
space. This is the first non-conventional assumption of our proposal\footnote{For an attempt to regard constants as solutions to an eigenvalue problem see, for example, Ref.~\cite{remo1}. Note that the focus of our paper is significantly different, once the structure of the Hilbert space and emphasis on non-comutativity are brought into play.}.

Next, we first assume that the various constants, $\kappa _{i}$, commute
with each other, so that their operators can be diagonalized simultaneously.
Then, there is an orthogonal basis of eigenvectors for all the $\hat{\kappa}%
_{i}$, satisfying: 
\begin{equation}
\langle \kappa _{1}^{\prime }...\kappa _{n}^{\prime }|\kappa _{1}...\kappa
_{n}\rangle =\prod_{i=1}^{n}\delta (\kappa _{i}^{\prime }-\kappa _{i}).
\label{braket}
\end{equation}%
We propose that being in an eigenstate of the $\hat{\kappa}_{i}$ entails not
only the observation of the values of the constants $\kappa _{i}$
corresponding to their eigenvalues, but also of the classical dynamics
defined by $\mathcal{H}(x,p;\kappa _{i})$ on the base space defined by $x,p$
and endowed with a Poisson bracket. 
This system is never classical overall; however, there is classical
behaviour, if we are in the eigenstate. Note that the constants are not
\textquotedblleft varying\textquotedblright\ in time or space, they can just
be observed to on take different values: each quantum outcome has different
fixed constants. We might think of this picture as a Hilbert space of
theories. As long as we stick to eigenstates we have a variation on the
theme of the \textquotedblleft multiverse\textquotedblright ~\cite{multi}, perhaps with a
more quantum flavour, but not substantially different in its implications.

We can ask what happens to the classical dynamics encoded in $\mathcal{H}%
(x,p;\kappa _{i})$ should we have a superposition of the $|\kappa
_{i}\rangle $? We postulate that we would then have a superposition of
classical phase spaces, each endowed with a different hamiltonian labelled
by the $\kappa _{i}$. That is, we would have a superposition of all the
solutions mapped from the initial conditions by the different evolutions.
This proposal has a definite flavour of the many-worlds interpretation of
quantum mechanics, but applied to the full set of classical solutions and
dynamics\footnote{For an earlier consideration of such superpositions see~\cite{sabine},
where the role of ``collapse'' of superpositions is also examined in the context of
variations of the speed of light and their implications for Lorentz invariance.}.

We represent the total hamiltonian of the system by 
\begin{equation}
\hat{\mathcal{H}}_{tot}(x,p)=\int d^{n}\kappa _{i}\,|\kappa _{i}\rangle
\langle \kappa _{i}|\otimes \mathcal{H}(x,p;\kappa _{i}).  \label{htot}
\end{equation}%
Here, the $\mathcal{H}(x;\kappa _{i})$ are (at least initially) classical,
and the Poisson bracket structure only applies to them. The $\hat{\mathcal{H}%
}_{tot}|\Psi \rangle $ represents the postulated superposition for the state 
$|\Psi \rangle \in \mathcal{H}$. The symbol $\otimes $ could represent the
usual bilinear tensor product or any other structure, possibly more complex.
Some notational simplification can be obtained by defining a composite
object 
\begin{equation}
x_{\Psi }(t)=|\Psi \rangle \otimes x=\int d\kappa _{i}\,\Psi (\kappa
_{i})|\kappa _{i}\rangle \otimes x(t;\kappa _{i}),  \label{composite}
\end{equation}%
representing the superposition of hamiltonian evolutions for the various
parameters $\kappa _{i}$, with amplitudes determined by the wavefunction: 
\begin{equation}
|\Psi \rangle =\int d\kappa _{i}\Psi (\kappa _{i})|\kappa _{i}\rangle .
\end{equation}%
Then, we can define 
\begin{eqnarray}
\dot{x}_{\Psi } &=&\{x_{\Psi },\hat{\mathcal{H}}_{tot}\},  \label{PB1} \\
\dot{p}_{\Psi } &=&\{p_{\Psi },\hat{\mathcal{H}}_{tot}\},  \label{PB2}
\end{eqnarray}%
as long as we assume that, by definition, all the quantum quantities to the
left of $\otimes $ act only on themselves as appropriate, and the Poisson
bracket only involves the classical quantities on the right of $\otimes $.
Indeed, the left-hand side of (\ref{PB1}) can then be written as: 
\begin{eqnarray}
\{|\Psi \rangle \otimes x,\hat{\mathcal{H}}_{tot}\} &=&\int d\kappa
_{i}|\kappa _{i}\rangle \langle \kappa _{i}|\Psi \rangle \otimes \{x,%
\mathcal{H}(x,p;\kappa _{i})\}  \notag \\
&=&\int d\kappa _{i}\Psi (\kappa _{i})|\kappa _{i}\rangle \otimes \dot{x}%
(t;\kappa _{i})  \notag \\
&=&|\Psi \rangle \otimes \dot{x}=\dot{x}_{\Psi },
\end{eqnarray}%
and likewise, for (\ref{PB2}).

As long as we live in an eigenstate of the constants' operator (which
assumes they commute), the proposal in this Section is very similar to the
treatment given to classical parameters in quantum information theory~\cite%
{Qinf}, where they are elevated to bras and kets with the property (\ref%
{braket}). The total hamiltonian is then written as (\ref{htot}), except
that in quantum information theory the $\mathcal{H}(x,p;\kappa _{i})\equiv 
\hat{\mathcal{H}}(\hat x,\hat p;\kappa _{i})$ already is a quantum
hamiltonian operator. Thus, at this stage, our proposal is quite standard,
as long as the $\Psi $ are eigenstates. Inevitable novelties arise though if
we now move on to assume that the $\hat{\kappa}_{i}$ do not commute.

\section{Non-commuting constants}

\label{noncmod}

Our model becomes significantly different if some of the constants do not
commute. In general, they might form an algebra: 
\begin{equation}
\lbrack \hat{\kappa}_{i},\kappa _{j}]=if_{ijk}\hat{\kappa}_{k},  \label{coms}
\end{equation}%
and later we will consider examples targeting particular ``constants'', for
example, $\hat{G}$, $\hat{c}$ and $\hat{\Lambda}$. This would introduce a
fundamental indeterminacy in their value, according to the Heisenberg
Uncertainty Principle. First consider the situation in which just two
constants are conjugates: 
\begin{equation}
\left[ \hat{\kappa}_{1},\hat{\kappa}_{2}\right] =i\mathfrak{h}  \label{com1}
\end{equation}%
where $\mathfrak{h}$ is a possibly dimensionful constant (the dimensions
will depend on the model), which may or may not be proportional to the usual
Planck constant. As is well known, following a purely kinematical argument,
we must have: 
\begin{equation}
\Delta ^{2}\kappa _{1}\Delta ^{2}\kappa _{2}\geq \frac{\mathfrak{h}^{2}}{4}.
\label{Hunc}
\end{equation}%
One might therefore think that the evolution would take place in the form of
a \textquotedblleft fuzzy\textquotedblright\ hamiltonian evolution in which
the parameters are undetermined, at least if the $|\Psi \rangle $
wavefunction is taken to be a coherent state.

However, this is not the case. It turns out that the evolution is
qualitatively different, and in fact there may not be any form of
conventional hamiltonian evolution at all. The key point is that we now have
a complementarity principle in operation: asking the system about the value
of one constant precludes asking questions about the other. Although it is
true that an eigenstate of, say, $\hat{\kappa}_{1}$ can be written as a
superposition of eigenstates of $\hat{\kappa}_{2}$, this superposition does 
\textit{not} represent a superposition of the simultaneous observation of
the original eigenvalue $\kappa _{1}$ and each of the eigenvalues of $\hat{%
\kappa}_{2}$. This is a trivial point, but it should be stressed. Therefore,
not only can we never be in an eigenstate of all the constants, but we can
also never be in a superposition of states with a well defined hamiltonian
in terms of the original constants. In terms of the original hamiltonian,
there can never be any hamiltonian evolution for the system, or even a
superposition thereof, because we cannot muster enough information to define
the classical hamiltonian, or even a superposition as envisaged in Section~%
\ref{simplestmod}.

In other words, the replacement 
\begin{equation}
\mathcal{H}(x,p;\kappa _{1},\kappa _{2})\rightarrow \mathcal{H}_{tot}(x,p;%
\hat{\kappa}_{1},\hat{\kappa}_{2})
\end{equation}%
may be done, but this can never be unravelled in the format (\ref{htot}),
resulting in a superposition of classical hamiltonian evolutions defined by $%
\mathcal{H}(x,p;\kappa _{1},\kappa _{2})$. At best, we might be able to
obtain, for example, something like: 
\begin{eqnarray}
\hat{\mathcal{H}}_{tot}(x,p)&=&\int d\kappa _{1}\,|\kappa _{1}\rangle
\langle \kappa _{1}|\otimes \mathcal{H}_{1}(x,p;\kappa _{1})  \notag \\
&+&\int d\kappa _{2}\,|\kappa _{2}\rangle \langle \kappa _{2}|\otimes 
\mathcal{H}_{2}(x,p;\kappa _{2}),  \notag
\end{eqnarray}%
if the classical (``proto'') hamiltonian can be split as 
\begin{equation}
\mathcal{H}(x,p;\kappa _{1},\kappa _{2})=\mathcal{H}_{1}(x,p;\kappa _{1})+%
\mathcal{H}_{2}(x,p;\kappa _{2}).
\end{equation}%
The lack of straightforward hamiltonian evolution would then be signalled by
the time derivatives of the states $x_{\Psi }$ and $p_{\Psi }$ not equalling
states in the original $|\Psi \rangle $ (cf. eqns.~(\ref{PB1}) and (\ref{PB2}%
)).

This does not mean that nothing can be said about the evolution of the
system, as we will show in the following Sections. The technical inputs of
the theory are

\begin{itemize}
\item[a.] The structure of the Hilbert space $\mathbb{H}$: its inner
product, and kinematic observables $\hat{\kappa}_{i}$).

\item[b.] The state $|\Psi \rangle \in \mathbb{H}$ in which we live.

\item[c.] The algebra of commutators (\ref{coms}).

\item[d.] The combined hamiltonian $\mathcal{H}_{tot}$ and its $\otimes $
structure.
\end{itemize}

Note that the $\otimes $ encodes how the quantum constants interact with the
classical hamiltonian. This can be non-linear and as complex as wanted.
Depending on the inputs above (point d in particular) we could conceivably
find that hamiltonian dynamics is impossible. However, it may also be
preserved in some form.


To clarify and summarise the proposal made in the last two sections: we have
elevated the constants of Nature to quantum operators, but these live in a
purely quantum kinematical Hilbert space. They do not have classical
counterparts, and their commutator does not result from a classical Poisson
bracket (although this could be explored and is left for future work; see~%
\cite{quantumlambda} for an instance of this approach regarding the quantum
cosmological constant). Concomitantly, the commutators of these constants do
not generate any time evolution for them: in this sense they are the genuine
constants.

Note also the difference with \cite{notime}, where the existence of a
dynamical hamiltonian for the constants is precluded by the fact that
localized time disintegrates, so that a hamiltonian structure is not
possible. Here, the kinematical nature of the constants' Hilbert space is a
choice, and it does blend in with a hamiltonian structure for $x$ and $p$.
The fact that the constants may not commute precludes in general the
existence of a hamiltonian evolution, at least in terms of the original
hamiltonian, since no state can summon enough information to specify the
hamiltonian. In some cases, a counterpart to a hamiltonian evolution can be
found, as we shall now see~\footnote{We stress that if a hamiltonian evolution cannot be found, then
concepts such as energy, pressure, partition functions, and equilibrium may simply not have
an operational meaning. 
This is not the case in the situation described in Section II, or if a counterpart of a hamiltonian evolution 
can be found when the constants do not commute, as will be assumed for the rest of this paper.}.

\section{An illustrative toy model}

\label{ho}

We will now apply our formalism to a metaphorical toy model -- not to be
taken literally as a physical model. It is intended only to illustrate our
proposal for the constants of Nature. Consider a harmonic oscillator with
action: 
\begin{equation}
S(x,\dot{x};m,k)=\int dt\left( \frac{p^{2}}{2m}-\frac{kx^{2}}{2}\right) ,
\end{equation}%
and hamiltonian: 
\begin{equation}
\mathcal{H}(x,p;m,k)=\frac{p^{2}}{2m}+\frac{kx^{2}}{2}.
\end{equation}%
Here, the spring constant, $k$, and the mass, $m$, play the role of
\textquotedblleft constants\textquotedblright\ of Nature. The proposal in
this paper is that $m$ and $k$ should be promoted to observables, $\hat{m}$
and $\hat{k}$, in an essentially kinematical Hilbert space, $\mathbb{H}$.

If these observables commute, and if we live in an eigenstate of the
constants' operator, the usual classical theory follows, with a treatment
given to its parameters reminiscent of quantum information theory. We would
then say that there are joint eigenstates $|k,m\rangle $ for which: 
\begin{equation}
\langle k^{\prime }m^{\prime }|k,m\rangle =\delta (k-k^{\prime })\delta
(m-m^{\prime }),
\end{equation}%
such that the total hamiltonian is given by 
\begin{equation}
\hat{\mathcal{H}}_{tot}(x,p)=\int dm\,dk\,|k,m\rangle \langle km|\otimes 
\mathcal{H}(x,p;m,k).
\end{equation}%
If the system lives in an eigenstate $|k,m\rangle $ then the dynamics
encoded in $\mathcal{H}(x,p;m,k)$ are observed. For a general superposition
of states $|\Psi \rangle $, we postulate a superposition of all the possible
dynamics and its solution $x_{\Psi }(t)$, with the amplitudes given by $%
\langle km|\Psi \rangle $.

Further quantum behaviour follows if $\hat{m}$ and $\hat{k}$ do not commute.
Then, we can never be in an eigenstate of both constants and, as stated
above, classical hamiltonian evolution in terms of the original hamiltonian
is generally not possible. A replacement for a hamiltonian evolution (or a
superposition thereof) might be found, however, depending on the assumptions
of the theory, as we now exemplify.

One example where some sort of hamiltonian evolution survives is when the $%
\otimes $ product is a simple bilinear tensor product. Let us consider,
furthermore, that: 
\begin{equation}
\hat{\mathcal{H}}_{tot}=\hat{\kappa}_{2}^{2}\otimes \frac{p^{2}}{2}+\hat{%
\kappa}_{1}^{2}\otimes \frac{x^{2}}{2},  \label{htot-osc}
\end{equation}%
that is, we write operators $\hat{m}$ and $\hat{k}$ in terms of variables: 
\begin{eqnarray}
\hat{\kappa}_{1}^{2} &=&\widehat{\frac{1}{m}} \\
\hat{\kappa}_{2}^{2} &=&\hat{k}
\end{eqnarray}%
with the assumption: 
\begin{equation}
\lbrack \hat{\kappa}_{1},\hat{\kappa}_{2}]=i\mathfrak{h}  \label{coms3}
\end{equation}%
(in this example dimensions imply $[\mathfrak{h}]=1/T$). As explained
before, $\hat{\mathcal{H}}$ is not to be seen as a hamiltonian for $\hat{%
\kappa}_{1}$ and $\hat{\kappa}_{2}$ (i.e., no Heisenberg equations for them
are to be generated): their Hilbert space is kinematical. In contrast, $x$
and $p$ are classical dynamical variables, with Poisson bracket: $\{x,p\}=1$.

Although no hamiltonian evolution is possible in terms of the original
constants, with such a simple $\otimes $ product we can diagonalize the
hamiltonian in $\hat{\kappa}_{1}$ and $\hat{\kappa}_{2}$ by introducing the
non-Hermitian operator $\hat{\alpha}$, such that: 
\begin{eqnarray}
\hat{\kappa}_{1}\otimes 1 &=&\frac{\hat{\alpha}+\hat{\alpha}^{\dagger }}{\sqrt 2}\otimes 
\sqrt{\frac{\mathfrak{h}p}{x}} \label{alpha1}\\
\hat{\kappa}_{2}\otimes 1 &=&\frac{\hat{\alpha}-\hat{\alpha}^{\dagger }}{{\sqrt 2} i}\otimes 
\sqrt{\frac{\mathfrak{h}x}{ p}}.\label{alpha2}
\end{eqnarray}%
Using the assumed bilinearity of $\otimes $ we can compute: 
\begin{equation}
\lbrack \hat{\alpha},\hat{\alpha}^{\dagger }]=1
\end{equation}%
and: 
\begin{equation}
\hat{\mathcal{H}}_{tot}=\frac{\hat{\alpha}\hat{\alpha}^{\dagger }+\hat{\alpha%
}^{\dagger }\hat{\alpha}}{2}\otimes \mathfrak{h}xp=\left( \hat{\alpha}%
^{\dagger }\hat{\alpha}+\frac{1}{2}\right) \otimes \mathfrak{h}xp.
\end{equation}%
We can therefore set up a Fock space representation for the Hilbert space
where $\hat{\kappa}_{1}$ and $\hat{\kappa}_{2}$ live, i.e. in terms of the
eigenstates $|n\rangle $ of $\hat{N}=\hat{\alpha}^{\dagger }\hat{\alpha}$.
In this representation the total hamiltonian is given by: 
\begin{equation}
\hat{\mathcal{H}}_{tot}=\sum_{n}|n\rangle \langle n|\otimes \mathfrak{h}%
\left( n+\frac{1}{2}\right) xp,  \label{Htot3}
\end{equation}%
so it is diagonal.

We find two interesting results. Firstly, we see that, although there is no
hamiltonian evolution in terms of the dynamics dependent on the original
constants (here $\kappa _{1}$ and $\kappa _{2}$), the system does have a
hamiltonian evolution in terms of another observable (here $\hat{N}$) quite
different from the original constants. Since any state in the Hilbert space
can be written as a superposition of the eigenstates of the new observable,
the most general evolution corresponds to a superposition of the new
hamiltonian evolutions. This includes the eigenstates of the original
constants or any coherent state written in terms of them. In terms of the
original constants, the evolution is therefore a quantum superposition of
hamiltonian evolutions. For an eigenstate of $\hat{\kappa _{1}}$ or ($\hat {%
\kappa _{2}}$) the amplitudes $\langle n|\kappa _{1}\rangle $ would be the
usual expressions in terms of Hermite polynomials. For a coherent state,
defined from $\hat{\alpha}|\alpha \rangle =\alpha |\alpha \rangle ,$ we
would have instead: 
\begin{equation}
\langle n|\alpha \rangle =e^{-\frac{|\alpha |^{2}}{2}}\frac{\alpha ^{n}}{%
\sqrt{n!}}.  \label{coherentcomps}
\end{equation}

Secondly, we see that each of the hamiltonian evolutions that the system
does admit is qualitatively different from the original one (possible only
if the original constants were classical numbers, or commuting quantum
operators and we selected one of their simultaneous eigenstates).
Specifically, from (\ref{Htot3}), we see that each classical hamiltonian
associated with $n$ is given by: 
\begin{equation}
\mathcal{H}_{n}(x,p;n)=\mathfrak{h}\left( n+\frac{1}{2}\right) xp.
\end{equation}%
The dynamical equations are therefore: 
\begin{eqnarray}
\dot{x} &=&\mathfrak{h}\left( n+\frac{1}{2}\right) x \\
\dot{p} &=&-\mathfrak{h}\left( n+\frac{1}{2}\right) p \\
\dot{\mathcal{H}}_{n} &=&0
\end{eqnarray}%
with solutions: 
\begin{eqnarray}
x &=&C\exp {\left[ \mathfrak{h}\left( n+\frac{1}{2}\right) t\right] } \\
p &=&C\exp {\left[ -\mathfrak{h}\left( n+\frac{1}{2}\right) t\right] }.
\end{eqnarray}%
Here $C$ is an integration constant related to the constant hamiltonian: 
\begin{equation}
\mathcal{H}_{n}=C^{2}\left( n+\frac{1}{2}\right) .
\end{equation}%
Closer examination shows that the qualitatively different dynamics can be
traced to the selection of positive frequencies states $|n\rangle$. States
with $n<0$ exist but are not normalizable. Excluding them demands $xp>0$, so
clearly the dynamics in $x,p$ phase space had to be significantly modified
(indeed by replacing the original harmonic oscillator by an inverted one
with a quantized instability constant).

\section{Direct cosmological applications}

\label{hocosmo} The toy model of the preceding Section turns out to be
formally very close to a real cosmological situation: a closed Friedmann
universe filled with radiation, which has the dynamics of a simple harmonic
oscillator\footnote{%
This is true for any perfect fluid with equation of state $P=(\gamma -1)\rho 
$, where $P$ is the pressure, in a closed Friedmann universe if we transform
the scale factor $a(\tau )$ to $y(\tau )=a^{(3\gamma -2)/2}$ where $\tau $
is the conformal time \cite{JDBOb}. This transformation shows that $y(\tau )$
satisfies the simple harmonic oscillator equation.}. Indeed, we have the
Friedmann equation for the metric scale factor $a(\tau )$, in conformal time 
$\tau $ with usual constant curvature parameter $K=0,$or $\pm 1$: 
\begin{equation}
a^{\prime 2}+Kc^{2}a^{2}=\frac{8\pi G}{3}M,  \label{fried1}
\end{equation}%
where $^{\prime }$ denotes a derivative with respect to $\tau $, so we have
the simple oscillator in conformal time: 
\begin{equation*}
a^{\prime \prime }=-Kc^{2}a.
\end{equation*}%
For a closed Friedmann radiation universe, we take 
\begin{eqnarray}
K &=&1, \\
M &=&\rho a^{4}.
\end{eqnarray}

This can be easily transposed into the model in Section IV. Note that for
our proposed constant quantization, it matters where we put the constants in
the initial classical hamiltonian -- as is often the case, equivalent
classical theories may lead to different quantum theories. In particular, it
is not indifferent for our procedure whether $G$ multiplies the matter
density, or divides the Einstein-Hilbert (EH) action, even though this is
equivalent classically. We will at first assume that the EH action is
multiplied by $c^{2}/G$, rather than having $G/c^{2}$ multiplying the matter
hamiltonian, before a transition to operators is imposed. With this choice
the hamiltonian has units of mass density. Other choices will be explored
later.

Specifically, let us assume a situation where we write the hamiltonian
constraint implied by (\ref{fried1}) as: 
\begin{equation}
\mathcal{H}(a,p;G,c)=G\frac{p^{2}}{2}+\frac{c^{2}}{G}\frac{a^{2}}{2}\approx 
\frac{4\pi M}{3}
\end{equation}%
where $\approx $ means on-shell and $p=a^{\prime }/G$ is implied by the
Hamilton equation: 
\begin{equation}
a^{\prime }=\{a,H\}=Gp,
\end{equation}%
with the usual Poisson bracket $\{a,p\}=1$. We can subject this classical
theory to our quantization of constants procedure, directly lifting results
from the previous Section (cf. Eq.\ref{htot-osc}), with the identifications: 
\begin{eqnarray}
\hat{\kappa}_{1}^{2} &=&\widehat{\frac{c^{2}}{G}},  \label{id1} \\
\hat{\kappa}_{2}^{2} &=&\hat{G}.  \label{id2}
\end{eqnarray}%
Thus, $\hat{\kappa}_{2}=\sqrt{\hat{G}}$ and $\kappa _{1}$ can be any linear
combination of $\hat{c}(\hat{G})^{-1/2}$ and $(\hat{G})^{-1/2}\hat{c}$, and
we can check that 
\begin{equation}
\lbrack \hat{c},\hat{G}]=2i\mathfrak{h}\hat{G}  \label{coms4}
\end{equation}%
enforces the required (\ref{coms3}). Concomitant with the hamiltonian having
units of a mass density, in this example, the quantity $\mathfrak{h}$
(playing the role of a dimensionally corrected Planck's constant of unknown
numerical value) has units of a speed.

We can therefore use the results of the previous Section to conclude that
the most general solution for the Friedmann metric scale factor is a
superposition of solutions of the form 
\begin{equation}
a=C\exp {\left[ \mathfrak{h}\left( n+\frac{1}{2}\right) \eta \right] ,}
\label{amilne}
\end{equation}%
with $n$ labelling the terms of the diagonalized hamiltonian (\ref{Htot3}),
and producing a discretized Hubble expansion rate. This is a superposition
of Milne universes, since they are exponential in conformal time, not proper
time, $t$, where $dt=ad\tau $. However, the dynamics behind it is very
different from that of the general relativistic Milne universe (for example,
we have $K=+1$, rather than $K=-1$). Note, for example, that the $a$
evolution can be obtained without $p$. We find for $p$, independently, that 
\begin{equation}
p=\frac{4\pi }{3}\frac{\rho a^{3}}{\mathfrak{h}(n+\frac{1}{2})},
\label{pmilne}
\end{equation}%
either directly from $\mathcal{H}_{n}\approx \frac{4\pi M}{3}$, or from the
Hamilton equation for $p$.

\section{More general equations of state}

\label{generalgamma} The previous result is very robust. It is valid
regardless of the matter equation of state. This can be checked via the
transformation to $y(\tau)$ mentioned above.  We present an alternative
method at the end of this Section, to which we refer the reader who wishes to 
skip the mathematics details. 
It can also be checked that the same result is obtained using
either conformal or proper time, dispelling obvious breaking of Lorentz
invariance. We start by explaining how to do more general calculations, deriving a 
general rule of thumb.

Note first that the algebraic manipulations leading to (\ref{Htot3}) do not
rely on the hamiltonian in $x$ and $p$ being that of a harmonic oscillator.
They rely only on the hamiltonian as a function of the quantized constants $%
\kappa _{1}$ and $\kappa _{2}$ being formally that of a quantum harmonic
oscillator. Therefore, in a more general setting we have the following
recipe. Start from a classical proto-hamiltonian of the form 
\begin{equation}
\mathcal{H}(x,p;\kappa _{1},\kappa _{2})=\frac{\kappa _{2}^{2}}{2}f(x,p)+%
\frac{\kappa _{1}^{2}}{2}g(x,p)+h(x,p).  \label{protoH}
\end{equation}%
Kinematically quantize its constants $\kappa _{1}$ and $\kappa _{2}$,
subject to (\ref{coms3}). Hence, (\ref{protoH}) is replaced by 
\begin{equation}
\hat{\mathcal{H}}_{tot}=\frac{\hat{\kappa}_{2}^{2}}{2}\otimes f(x,p)+\frac{%
\hat{\kappa}_{1}^{2}}{2}\otimes g(x,p)+h(x,p).
\end{equation}%
As in the calculation leading to (\ref{Htot3}), this can be diagonalized to 
\begin{equation}
\hat{\mathcal{H}}_{tot}=\sum_{n}|n\rangle \langle n|\otimes \left( \mathfrak{%
h}\left( n+\frac{1}{2}\right) \sqrt{fg}+h\right) ,  \label{Htot4}
\end{equation}%
where the last term (in $h$) follows from 
\begin{equation}
\mathbb{1}=\sum_{n}|n\rangle \langle n|.
\end{equation}%
Therefore, \textit{as a rule of thumb}, the hamiltonian can be diagonalized
into terms $\mathcal{H}_{n}$, (labelled by a discrete observable $n$
replacing $\kappa _{i}$), each containing a term multiplying $\mathfrak{h}%
(n+1/2)$ and \textit{the geometrical mean of the original factors
multiplying $\kappa _{1}^{2}$ and $\kappa _{2}^{2}$}, to which one must add
any terms left outside the non-commuting variables (here denoted by $h$). In
addition, any constraint valid for the original hamiltonian (e.g. $\mathcal{H%
}\approx 0$) is now applicable to each of the diagonal components (e.g. $%
\mathcal{H}_{n}\approx 0$).

Equipped with this rule of thumb, we can now explore more general settings.
Let us work with proper time, $t$, for a change (the calculation is almost
identical with conformal time) and take the proto-hamiltonian: 
\begin{equation}
\mathcal{H}=\frac{Gp^{2}}{2}+\frac{c^{2}}{2G}-\frac{4\pi }{3}\rho a^{2},
\label{protoH1}
\end{equation}%
with the on-shell constraint $\mathcal{H}\approx 0$ and the separate
condition: 
\begin{equation}
\rho =\frac{M}{a^{3\gamma }},  \label{rhogamma}
\end{equation}%
(we recall that the equation of state is given by $P=(\gamma -1)\rho $,
where $P$ is the pressure and $\gamma $ is constant). In (\ref{protoH1}), $%
\rho $ is given by (\ref{rhogamma}) and is to be seen as a function of $a$.
It can be checked that this leads to the Friedmann and Raychaudhuri
equations via Hamilton's equations and/or the constraint (note that $p=\dot{a%
}/G$, where the dot denotes derivative with respect to proper time and $K=+1$%
). Upon quantization with the same identifications and assumptions as in
Section~\ref{hocosmo} (i.e. (\ref{id1}) and (\ref{id2}), with commutator (%
\ref{coms4}) enforcing (\ref{coms3})), we therefore replace $\mathcal{H}$
with 
\begin{equation}
\hat{\mathcal{H}}_{tot}=\frac{\hat{\kappa}_{2}^{2}}{2}\otimes p^{2}+\frac{%
\hat{\kappa}_{1}^{2}}{2}\otimes 1-\frac{4\pi }{3}\rho a^{2},
\end{equation}%
and this diagonalizes to 
\begin{equation}
\hat{\mathcal{H}}_{tot}=\sum_{n}|n\rangle \langle n|\otimes \left( \mathfrak{%
h}\left( n+\frac{1}{2}\right) p-\frac{4\pi }{3}\rho a^{2}\right) .
\label{Htot4a}
\end{equation}%
Each $\mathcal{H}_{n}$ term in the sum leads to 
\begin{equation}
\dot{a}=\{a,\mathcal{H}_{n}\}=\mathfrak{h}\left( n+\frac{1}{2}\right)
\end{equation}%
so that: 
\begin{eqnarray}
a &=&\mathfrak{h}(n+\frac{1}{2})t, \\
p &=&\frac{4\pi }{3}\frac{\rho a^{2}}{\mathfrak{h}(n+\frac{1}{2})},
\end{eqnarray}%
which is nothing but (\ref{amilne}) and (\ref{pmilne}) in terms of proper
time instead of conformal time. As announced  the results derived in the previous Section 
for radiation {\it do not in fact depend on
the equation of state}. They can also be equally obtained with proper and
conformal time. Indeed, as the first Hamilton equation for (\ref{Htot4})
shows ($\dot{a}=\{a,\mathcal{H}_{n}\}$), the expansion factor $a$ is blind
to matter, or put in another way, matter does not appear to gravitate. The
quantized Hubble constant is given by: 
\begin{equation}
H=\frac{\dot{a}}{a}=\frac{\mathfrak{h}(n+\frac{1}{2})}{t}.
\end{equation}

\section{Further models based on the same algebra}

\label{Lambda-mods}

Bearing in mind the rule of thumb derived at the start of Section~\ref%
{generalgamma}, we can now explore the effects of imposing a similar algebra
on other constants of Nature in the same setting. We begin by introducing
the cosmological constant, $\Lambda $.

\subsection{Models involving $\Lambda $}

We could have started with a proto-hamiltonian, 
\begin{equation}
\mathcal{H}=\frac{Gp^{2}}{2}+\frac{Kc^{2}}{2G}-\frac{\Lambda c^{2}a^{2}}{6G}-%
\frac{4\pi }{3}\rho a^{2},  \label{protoH2}
\end{equation}%
in order to accommodate a general curvature $K$ and a geometrical
cosmological constant (i.e. a cosmological term with units $1/L^{2}$ arising
from the geometrical part of the action). Then, with the same assumptions
(specifically, Eqns.~\ref{id1}, \ref{id2}, \ref{coms4}), this would
translate into: 
\begin{equation}
\hat{\mathcal{H}}_{tot}=\frac{\hat{\kappa}_{2}^{2}}{2}\otimes p^{2}+\frac{%
\hat{\kappa}_{1}^{2}}{2}\otimes \left( K-\frac{\Lambda }{3}a^{2}\right) -%
\frac{4\pi }{3}\rho a^{2}.
\end{equation}%
So long as 
\begin{equation}
K-\frac{\Lambda }{3}a^{2}>0,
\end{equation}%
the quantization and diagonalization can proceed as before (if this
condition is violated we would be quantizing an inverted harmonic
oscillator; we leave this for future work). Then, applying the rule of thumb
given in Section~\ref{generalgamma}, we obtain, 
\begin{eqnarray}
\hat{\mathcal{H}}_{tot} &=&\sum_{n}|n\rangle \langle n|\otimes \mathcal{H}%
_{n}  \label{Htot5} \\
\mathcal{H}_{n} &=&\mathfrak{h}\left( n+\frac{1}{2}\right) p\sqrt{K-\frac{%
\Lambda }{3}a^{2}}-\frac{4\pi }{3}\rho a^{2}.
\end{eqnarray}%
For each $n$, the first Hamilton equation gives: 
\begin{equation}
\dot{a}=\{a,\mathcal{H}_{n}\}=\mathfrak{h}\left( n+\frac{1}{2}\right) \sqrt{%
K-\frac{\Lambda }{3}a^{2}}.  \label{hameq2}
\end{equation}%
This can be easily integrated for the various cases $K=0,\pm 1$ and $\Lambda
>$ or $\leq 0$.

If $K=0$ and $\Lambda <0$, for all matter equations of state we find
exponential expansion: 
\begin{eqnarray}
a &=&C\exp {\left( Ht\right) } \\
H &=&\mathfrak{h}\left( n+\frac{1}{2}\right) \sqrt{\frac{|\Lambda |}{3}}
\end{eqnarray}%
with a quantized Hubble constant $H$. We see that even though the matter $%
\rho $ dominates at early times, $\Lambda $ controls the dynamics of $a$.
Matter, however, still dominates the conjugate momentum: 
\begin{equation}
p=\frac{4\pi }{3}\frac{\rho a}{\sqrt{\frac{|\Lambda |}{3}\mathfrak{h}(n+%
\frac{1}{2})}}.
\end{equation}%
Likewise, it can be checked that $K=-1$ and $\Lambda >0$ lead to AdS-type
expansion 
\begin{equation}
a=A\sin \left( \mathfrak{h}\left( n+\frac{1}{2}\right) \sqrt{\frac{|\Lambda |%
}{3}}\right)
\end{equation}%
with $p$ obtained from $\mathcal{H}_{n}=0$.

We could continue listing solutions, but by now a pattern has emerged, which
can be understood. Note that (\ref{hameq2}) can be seen as the expanding
branch of the alternative effective Friedmann equation: 
\begin{equation}
H^{2}=\left( \frac{\dot{a}}{a}\right) ^{2}=\mathfrak{h}^{2}\left( n+\frac{1}{%
2}\right) ^{2}\left( \frac{K}{a^{2}}-\frac{\Lambda }{3}\right) ,
\end{equation}%
(the analogy is only partial and fails to capture the behaviour of the
conjugate momentum $p$). In passing, we note that the contracting branch
would result from the negative energy spectrum (the \textquotedblleft Dirac
sea\textquotedblright ), and will be studied elsewhere (its wave functions
are not normalizable). Therefore, at least regarding $a$, the dynamics of
the diagonalized hamiltonian is similar to GR but replacing the original
non-commuting constants $c$ and $G$ by the effective constants: 
\begin{eqnarray}
c &\rightarrow &\mathfrak{h}\left( n+\frac{1}{2}\right) ,  \label{ceff} \\
G &\rightarrow &0;  \label{Geff}
\end{eqnarray}%
that is, $G$ is switched off and $c$ becomes quantized. In addition, the
modified dynamics effects sign changes in the other parameters: 
\begin{eqnarray}
K &\rightarrow &-K \\
\Lambda &\rightarrow &-\Lambda .
\end{eqnarray}%
This explains the counter-intuitive pattern obtained in our solutions. We
see why, even though $\rho $ dominates the curvature and $\Lambda $ terms,
the expansion is blind to it in the diagonalized dynamics: $G$ has switched
off and matter is a mere spectator because it does not gravitate.
Unsurprisingly, all our solutions for $a$ are also independent of the matter
equation of state. We see also why $K=1$ leads to Milne expansion: $K$
appears with opposite sign in the effective dynamics. Likewise for
exponential expansion with $K=0$ and $\Lambda <0$ and AdS expansion for $K=1$
and $\Lambda >0$. The Hubble parameters appear quantized because the
conversion constant $c$ is quantized.

\subsection{Complementary $G$ and $\Lambda$}

\label{candL}

One of the solutions above can be reproduced from a fundamentally different
standpoint if $G$ and $\Lambda $ are promoted to operators that do not
commute among themselves, but which do commute with $c$ and all the other
constants of the theory. Assume that $\Lambda $ is forced to be negative
(how a formalism can be developed to determine the sign of a constant will
be presented later in this paper), and that: 
\begin{eqnarray}
\hat{\kappa}_{1}^{2} &=&\widehat{\frac{|\Lambda |}{G}},  \label{id3} \\
\hat{\kappa}_{2}^{2} &=&\hat{G},  \label{id4}
\end{eqnarray}%
with (\ref{coms3}) assumed (in this case dimensionally $[\mathfrak{h}]=1/L$%
). Then, if $K=0$, the proto-hamiltonian (\ref{protoH2}) now translates to: 
\begin{equation}
\hat{\mathcal{H}}_{tot}=\frac{\hat{\kappa}_{2}^{2}}{2}\otimes p^{2}+\frac{%
\hat{\kappa}_{1}^{2}}{2}\otimes \frac{c^{2}a^{2}}{3}-\frac{4\pi }{3}\rho
a^{2},
\end{equation}%
and diagonalizing into pieces with 
\begin{equation}
\mathcal{H}_{n}=\mathfrak{h}\left( n+\frac{1}{2}\right) \frac{pca}{\sqrt{3}}-%
\frac{4\pi }{3}\rho a^{2}.
\end{equation}%
For all $\gamma $, each of these leads to: 
\begin{equation}
\dot{a}=\{a,\mathcal{H}_{n}\}=\mathfrak{h}\left( n+\frac{1}{2}\right) \frac{%
ac}{\sqrt{3}}.  \label{hameq3}
\end{equation}%
and to a solution: 
\begin{equation}
a=C\exp {\left( \frac{\mathfrak{h}c}{\sqrt{3}}\left( n+\frac{1}{2}\right)
t\right) }.
\end{equation}%
Even though the solution is very similar to that above, the setting is
different. Instead of (\ref{ceff}) and (\ref{Geff}) we have induced the
effective changes: 
\begin{eqnarray}
\Lambda &\rightarrow &\mathfrak{h}^{2}\left( n+\frac{1}{2}\right) ^{2}, \\
G &\rightarrow &0.
\end{eqnarray}%
That is, we have switched off $G$, as before, but now quantized $\Lambda $
instead of $c$. In addition, $\Lambda $ effectively acts as if its sign has
been reversed. Including $K$ here is also very different, since the quantum
harmonic oscillator will then acquire an interaction term.

\subsection{Models leading to static universes}

\label{static}

It could be that the classical theory from which we start is not general
relativity, but some extension or modification of it, adding new parameters
that can be constrained by observations. The prospect of quantizing these
parameters, however, might make them very relevant in the early universe,
even if irrelevant to the classical dynamics at late times. We illustrate
this point with an example.

For simplicity, consider a Friedmann model with $K=\Lambda =0$ and
proto-hamiltonian containing some parameter $\zeta $: 
\begin{equation}
\mathcal{H}=\frac{Gp^{2}}{2}+\frac{\zeta }{2p^{2}}-\frac{4\pi }{3}\rho
a^{2}\approx 0.  \label{protoH3}
\end{equation}%
Even at this stage, the first Hamilton equation leads to a novelty: 
\begin{equation}
\dot{a}=\{a,\mathcal{H}\}=Gp-\frac{\zeta }{p^{3}},
\end{equation}%
so that the Friedmann equation is modified even classically. Therefore a
dimensionless combination involving $\zeta $ must be small, but this does
not mean the implications cannot be large when its quantization is relevant.

Indeed, if we now quantize as in the previous examples, but with
identifications: 
\begin{eqnarray}
\hat{\kappa}_{1}^{2} &=&\hat{\zeta}  \label{id5} \\
\hat{\kappa}_{2}^{2} &=&\hat{G}  \label{id6}
\end{eqnarray}%
we get the quantum hamiltonian: 
\begin{equation}
\hat{\mathcal{H}}_{tot}=\frac{\hat{\kappa}_{2}^{2}}{2}\otimes p^{2}+\frac{%
\hat{\kappa}_{1}^{2}}{2}\otimes \frac{1}{p^{2}}-\frac{4\pi }{3}\rho a^{2},
\end{equation}%
which results in diagonalized pieces: 
\begin{equation}
\mathcal{H}_{n}=\mathfrak{h}\left( n+\frac{1}{2}\right) -\frac{4\pi }{3}\rho
a^{2}.
\end{equation}%
Therefore, the Hamilton equation for $a$ for each of these pieces gives: 
\begin{equation}
\dot{a}=0
\end{equation}%
with the hamiltonian constraint, $\mathcal{H}_{n}\approx 0$, fixing: 
\begin{equation}
a_{n}=\frac{4\pi }{3}\frac{\rho }{\left( n+\frac{1}{2}\right) }.
\end{equation}%
The fact that $G$ and $\zeta $ may become complementary variables at early
times therefore makes $\zeta $ relevant, no matter how small it may be
classically. Specifically, we need the wave function to be a coherent state
centered around $\zeta =0$ and the current value of $G$. Early on, this
decomposes into a superposition of Fock states (as in Eq.\ref{coherentcomps}%
), each corresponding to a static universe with a different constant
expansion factor, and thus containing different amounts of matter in the
same comoving region.


\section{Switches and the algebra of $SU(2)$}
\label{su2} Another class of applications results from taking three
constants, considering a piece of binary information about them (for example
their sign, or whether they are switched off to zero, or switched on at a
fixed value), and place this information in the fundamental representation
of $SU(2)$. This binary information can only then be known about one of the
constants, with the others just working as complementary discrete variables.
The general framework is most easily illustrated by two extreme cases: when
we are concerned with the sign of two constants; and when we use the algebra
to switch all three on or off. Mixed cases will be explained at the end of
this Section.

\subsection{The signs of the constants}

Consider a proto-hamiltonian made up of four terms: 
\begin{equation}
\mathcal{H}=f+g+h+r  \label{protoH0b}
\end{equation}%
for which we will explore the sign of the first three. Here, all terms are
functions of generic $x$ and $p$, as well as constants $\kappa _{i}$ (which
we assume can be treated classically or as eigenvalues of operators for
which the universe is in an eigenstate). If the constants $\kappa _{i}$
appear multiplicatively in these terms, we are effectively exploring their
sign.

As an example, we could revisit the toy model of Section~\ref{ho}, add an
aharmonic term to it, and set: 
\begin{eqnarray}
f &=&\frac{p^{2}}{2m},  \notag \\
g &=&\frac{kx^{2}}{2},  \notag \\
h &=&\lambda x^{4},  \notag \\
r &=&0.  \notag
\end{eqnarray}%
Our models here would therefore concern the sign of $m$, $k$ and $\lambda $;
that is, whether we have a ghost or not, if the oscillator is inverted or
not, and the sign of the quartic term.

As before, the idea is to replace the proto-hamiltonian by a quantum
hamiltonian in which the aspect under consideration is promoted to an
operator in a kinematical Hilbert space. Specifically, we replace (\ref%
{protoH0b}) by 
\begin{equation}
\hat{\mathcal{H}}_{tot}=\hat{\sigma}_{1}\otimes f+\hat{\sigma}_{2}\otimes g+%
\hat{\sigma}_{3}\otimes h+r,  \label{htotb1}
\end{equation}%
where $\sigma _{i}$ are Pauli matrices. Thus the sign of a given term is
fixed if we are in an eigenstate of its corresponding $\sigma _{i}$; however
the sign of more than one of the terms can never be known, due to the laws
of quantum mechanics. As a result the original hamiltonian dynamics cannot
be used to propagate the system.

Nevertheless, as before, if the $\otimes $ product is trivial, the
hamiltonian can be diagonalized. The system then generically evolves as a
superposition of alternative hamiltonians. Diagonalization of (\ref{htotb1})
leads to 
\begin{equation}
\hat{\mathcal{H}}_{tot}=\sum_{S=\pm }|S\rangle \langle S|\otimes \left( r+S%
\sqrt{f^{2}+g^{2}+h^{2}}\right) ,
\end{equation}%
with a single \textquotedblleft sign\textquotedblright\ $S=\pm 1$. If one of
the three terms dominates, $S$ corresponds to the choice of sign for that
term in the original hamiltonian. If two or three of these terms are
comparable we must replace the original hamiltonian by the root-mean-square
(RMS) average of the three terms before taking the $\pm $ sign associated
with $S$.

\subsection{Switching the constants on and off}

A similar model can be set up by shifting the Pauli matrices so that their
eigenvalues are either 0 or 1. The eigenstates are then associated with
switching on and off the corresponding term (or constant, if it appears
multiplicatively in the term). If we were to apply this to all three terms,
we would have: 
\begin{equation}
\hat{\mathcal{H}}_{tot}=\frac{\hat{\sigma}_{1}+\mathbb{1}}{2}\otimes f+\frac{%
\hat{\sigma}_{2}+\mathbb{1}}{2}\otimes g+\frac{\hat{\sigma}_{3}+\mathbb{1}}{2%
}\otimes h+r.  \notag
\end{equation}%
Diagonalization would then result in: 
\begin{equation}
\hat{\mathcal{H}}_{tot}=\sum_{S=\pm }|S\rangle \langle S|\otimes \left( r+%
\frac{f+g+h}{2}+S\sqrt{\frac{f^{2}+g^{2}+h^{2}}{4}}\right) .  \notag
\end{equation}%
Again, if one of the three terms dominates then the hamiltonian does for
that term what would be expected by identifying $S$ and the eigenvalue for
that term ($S=1$ switches on, $S=-1$ switches off). If the terms are
comparable instead, then a suitable average is made, based on the RMS.

Given the patterns found in the two extreme cases above, we can guess the
result for hybrid cases, where we are concerned about the sign of some terms
but the on/off nature of others. For example 
\begin{equation}
\hat{\mathcal{H}}_{tot}=\hat{\sigma}_{1}\otimes f+\frac{\hat{\sigma}_{2}+%
\mathbb{1}}{2}\otimes g+\frac{\hat{\sigma}_{3}+\mathbb{1}}{2}\otimes h+r. 
\notag
\end{equation}%
would result in: 
\begin{equation}
\hat{\mathcal{H}}_{tot}=\sum_{S=\pm }|S\rangle \langle S|\otimes \left( r+%
\frac{g+h}{2}+S\sqrt{f^{2}+\frac{g^{2}+h^{2}}{4}}\right) .  \notag
\end{equation}

\section{Cosmology and the sign of the constants}

\label{su2cosmo}

We now consider a cosmological application of the formalism developed in
Section~\ref{su2}, starting from the proto-hamiltonian: 
\begin{equation}
\mathcal{H}=\frac{p^{2}c^{2}}{4}+K-\frac{\Lambda a^{2}}{3}-\frac{8\pi G}{%
3c^{2}}\rho a^{2}.  \label{protoH1b}
\end{equation}%
It can be readily proved that the hamiltonian constraint, $\mathcal{H}%
\approx 0$, implies the Friedmann equation (note that with this hamiltonian $%
\dot{a}=\{a,\mathcal{H}\}=pc^{2}/2$). With this choice of \textquotedblleft
where to put the constants\textquotedblright\ the hamiltonian has units $[%
\mathcal{H}]=1/L^{2}$. We subject this model to the formalism of Section~\ref%
{su2} with choices and notation: 
\begin{eqnarray}
f &=&K \\
g &=&-\frac{\Lambda a^{2}}{3}=-\lambda a^{2} \\
h &=&-\frac{8\pi G}{3c^{2}}\rho a^{2}=-\frac{m}{a^{2}} \\
r &=&\frac{p^{2}c^{2}}{4}
\end{eqnarray}%
where, for simplicity, we have assumed a radiation equation of state, $%
\gamma =4/3$ (but what follows generalizes straightforwardly to other $%
\gamma $). We can now consider what happens for the various cases in which
we use $SU(2)$ to probe the sign or the on/off nature of the various terms.
For each case we have two types of dynamics labelled by the eigenvalue $S$.
In some cases no qualitative novelties were found. Below we highlight those
where significantly different behaviours emerge.

\subsection{On and off switches for $K$, $\Lambda $ and $G$}

Imagine a situation in which we fix the values of $K$, $\Lambda $ (which may
be positive or negative) and $G$ (assumed positive for simplicity), and
insert the information on whether they are switched off or on (set to these
values) into the algebra of $SU(2)$. Then, the hamiltonian constraint for
each $\mathcal{H}_{S}$ in 
\begin{equation}
\hat{\mathcal{H}}_{tot}=\sum_{S=\pm }|S\rangle \langle S|\otimes \mathcal{H}%
_{S}
\end{equation}%
implies the Friedmann equations for each eigenstate labelled by $S=\pm 1$: 
\begin{equation}
H^{2}=\frac{m}{2a^{4}}-\frac{K}{2a^{2}}+\frac{\lambda }{2}+S\sqrt{\frac{m^{2}%
}{4a^{8}}+\frac{1}{4a^{4}}+\frac{\lambda ^{2}}{4}}.
\end{equation}%
The presence of either a bounce or a turnaround is immediately revealed by
the zeros of the right-hand side, should we choose the $S=-1$ branch: 
\begin{equation}
a^{2}=\frac{\lambda }{2K}\left( 1\pm \sqrt{1-\frac{4}{m\lambda }}\right) .
\label{roots}
\end{equation}%
Real roots require $m\lambda \geq 4$, so with $m>0$ (i.e. positive $G$) we
must also have $\Lambda >0$. Since $a^{2}$ must be positive (\ref{roots})
then implies $K=1$. By drawing the effective potential (defined from $\dot{a}%
^{2}+V_{eff}=0$), we see that indeed the universe oscillates between a
bounce and a turnaround in this branch. This can be inferred from Fig.~\ref%
{veff1}, where we depicted $V_{eff}$ for $K=1$, $m=4$ and $\lambda =4$. The
universe expands and contracts in the region where $V_{eff}$ is negative
(with kinetic terms given by $\dot{a}^{2}=-V_{eff}$), with a bounce
(smallest root) and a turnaround (largest root) where $V_{eff}=0$. 
\begin{figure}[h]
\begin{center}
\scalebox{0.5}{\includegraphics{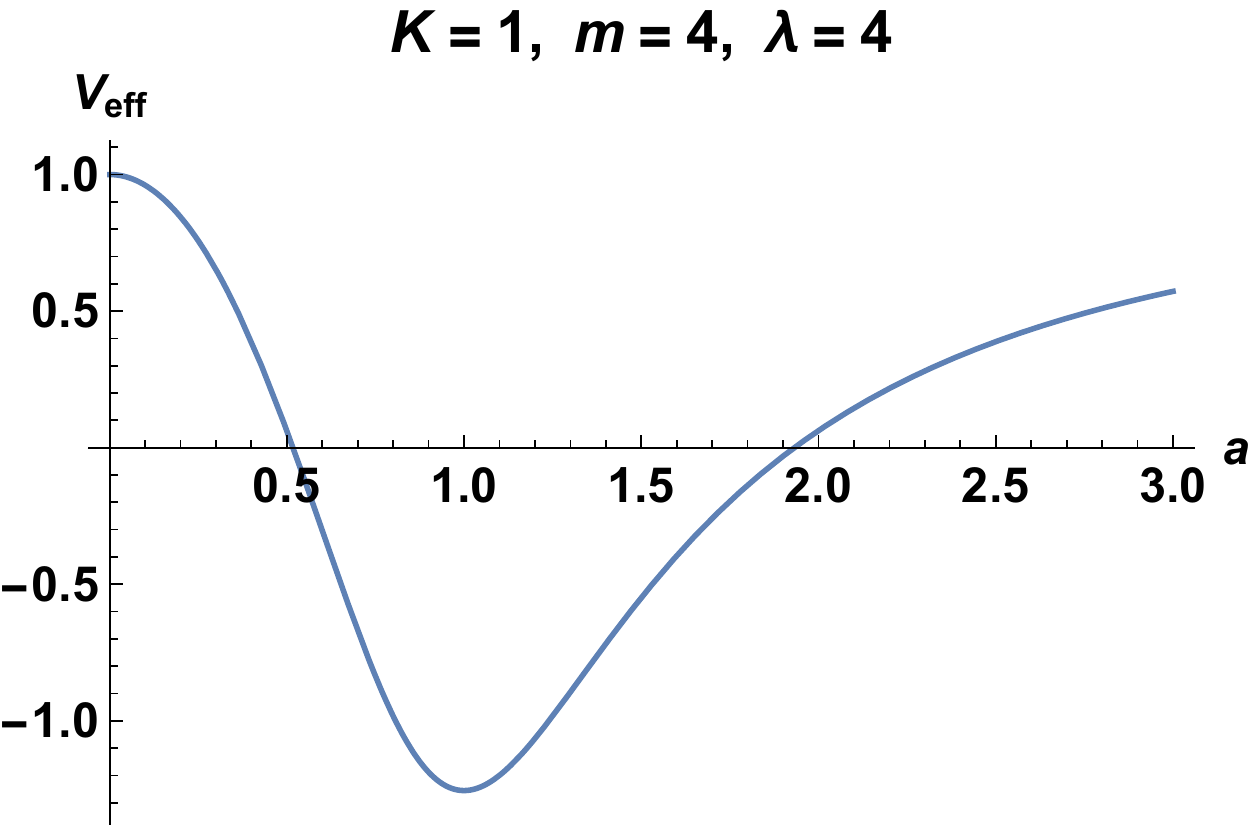}}
\end{center}
\caption{The effective potential for $K=1$, $m=4$ and $\protect\lambda =4$.
The universe expands and contracts in the region where $V_{eff}<0$ (with $%
\dot{a}^{2}=-V_{eff}$), with a bounce and a turnaround where $V_{eff}=0$. }
\label{veff1}
\end{figure}

\begin{figure}[h]
\begin{center}
\scalebox{0.5}{\includegraphics{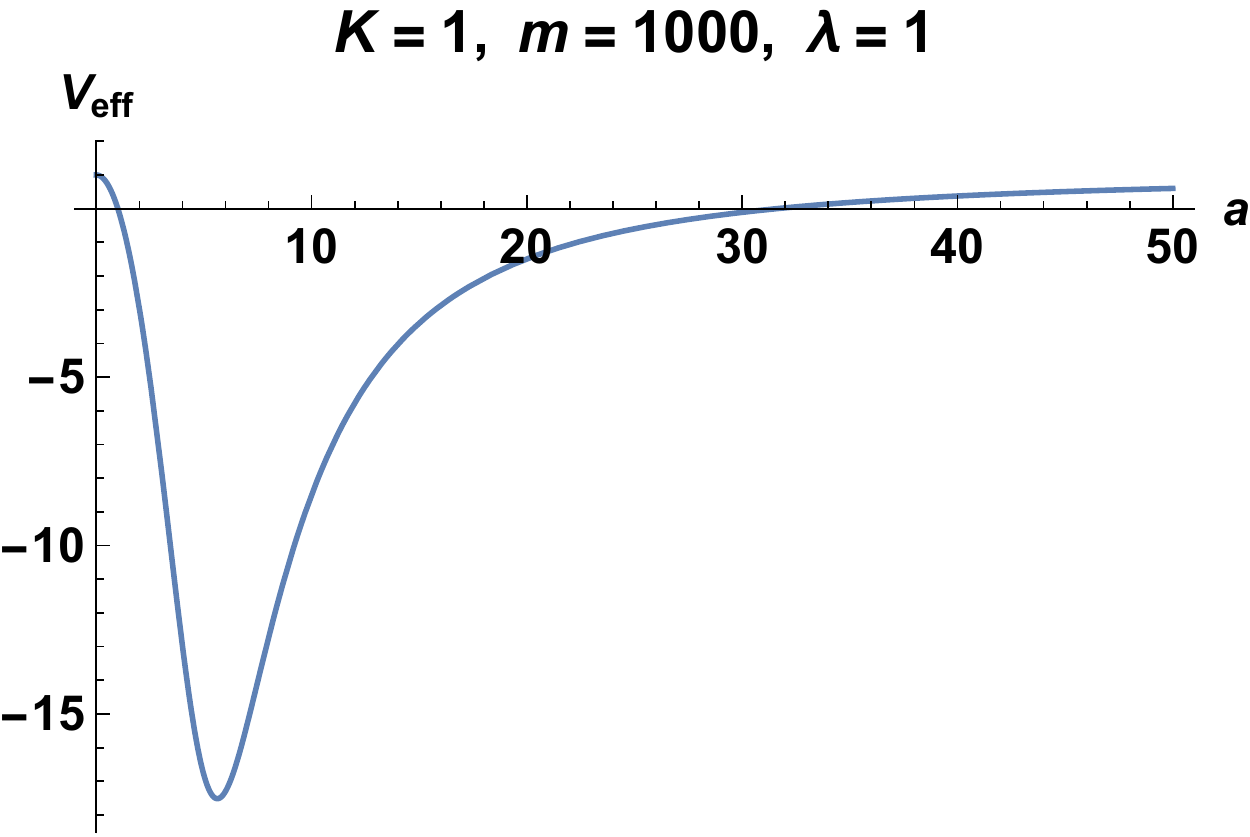}}
\end{center}
\caption{The effective potential for $K=1$, $m=1000$ and $\protect\lambda =4$%
. As we see by dialing $m$ and $\protect\lambda $ we can make the bounce as
close to zero as required, and the expansion cycle as large as desired.}
\label{veff2}
\end{figure}

\begin{figure}[h]
\begin{center}
\scalebox{0.5}{\includegraphics{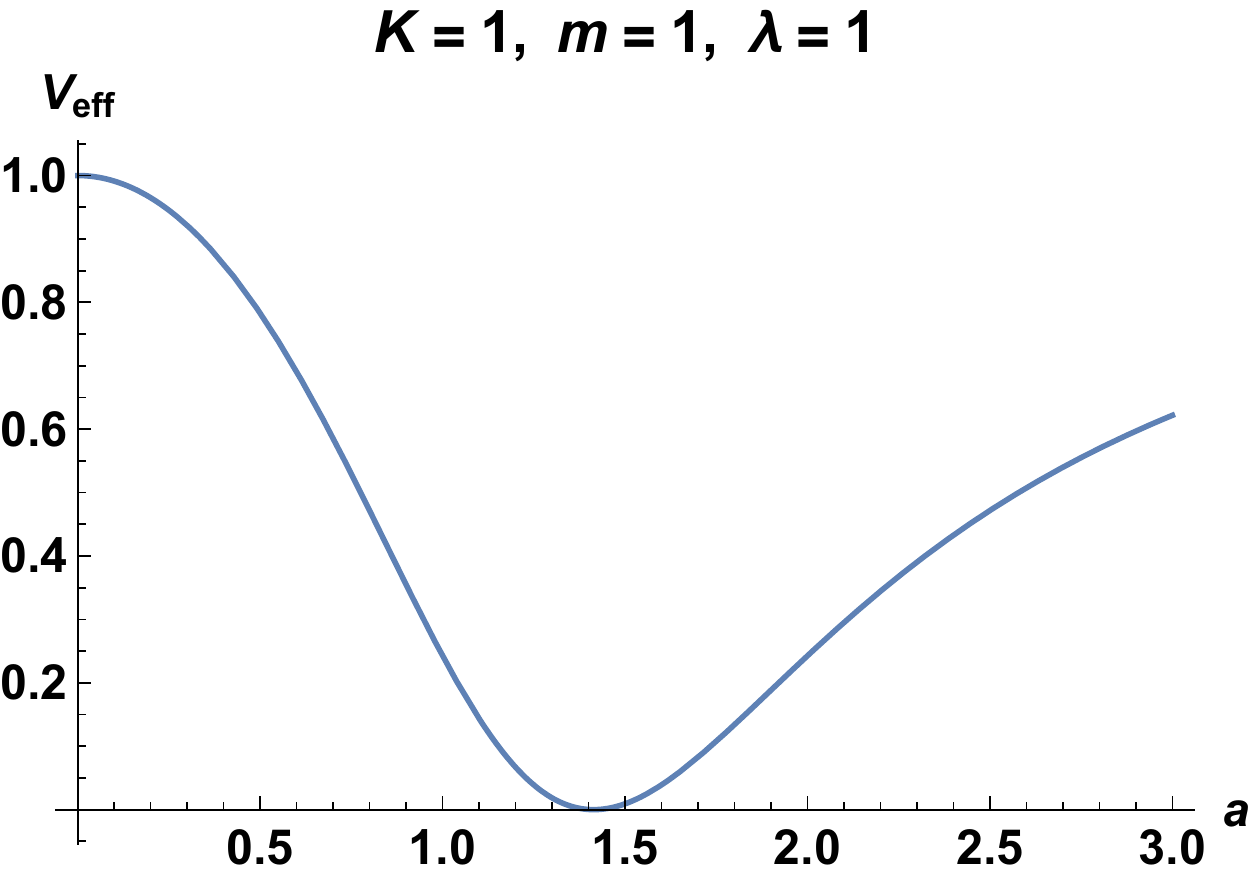}}
\end{center}
\caption{The effective potential for extremal case $K=1$, $m=1$ and $\protect%
\lambda =4$. As the picture shows, a stable static Universe is found in this
case. }
\label{veff3}
\end{figure}

As illustrated in Fig.~\ref{veff2}, and as can be read off from (\ref{roots}%
), a variety of scenarios can be arranged by dialling $m$ and $\Lambda $.
Specifically, by changing $m$ and $\lambda $ we can make the bounce as close
to zero as required, and the expansion cycle as large as wanted. 
On the other extreme, by setting $\lambda m=4$, we obtain a static universe
with: 
\begin{equation}
a=\sqrt{\frac{\Lambda }{6}}.
\end{equation}%
Not only are these conditions different from the usual Einstein static
universe \cite{static}, but also, as expected from the behaviour when $%
\lambda m>4$, we see that such a static universe is \textit{stable}. The
effective potential is depicted in Fig.~\ref{veff3} for this case. 

\subsection{A single sign switch and two on/off switches}

The only other situation where qualitative novelties are found is when one
constant or term is subject to a sign switch and the other two have an
on/off switch within the algebra of $SU(2)$. Three different cases can be
listed:

\subsubsection{Case I}

\begin{figure}[h]
\begin{center}
\scalebox{0.5}{\includegraphics{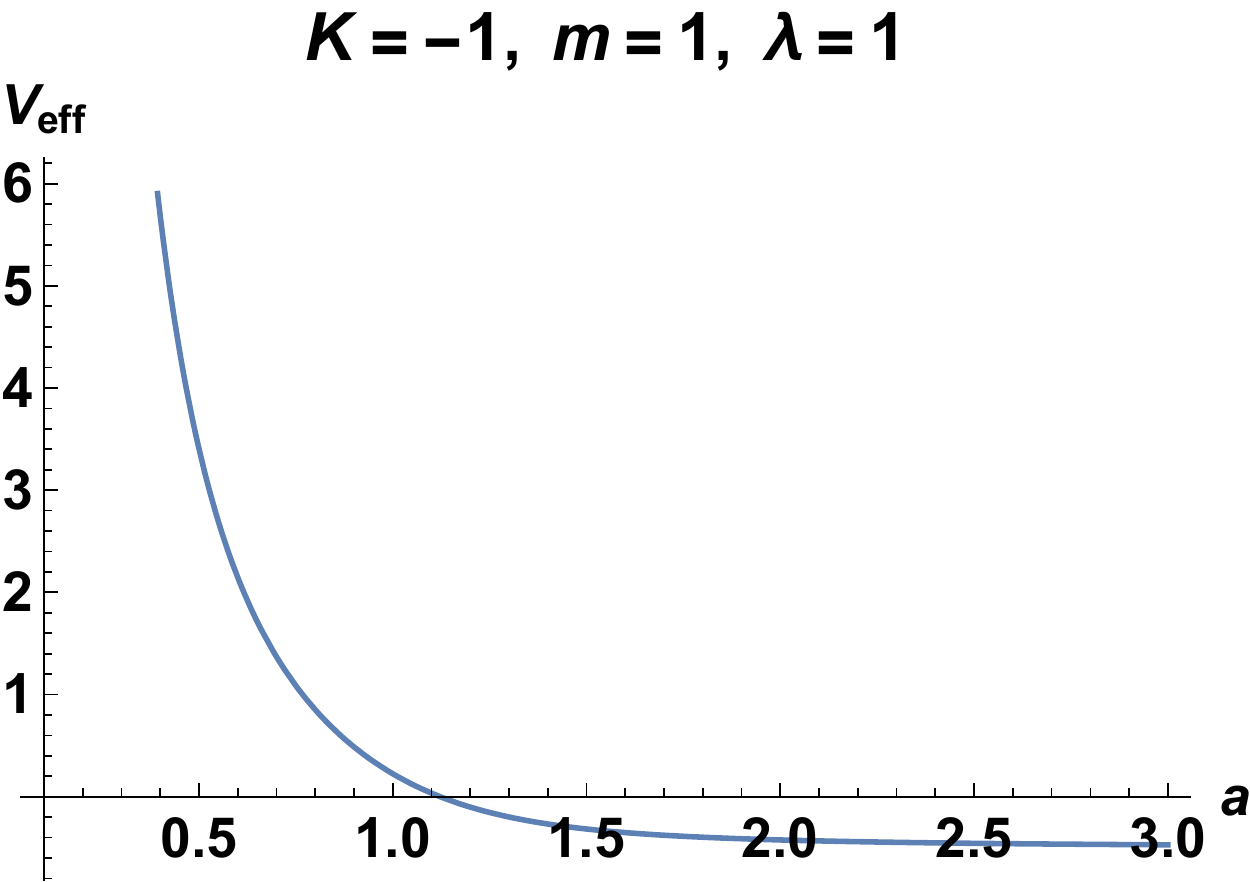}}
\end{center}
\caption{The effective potential when we query the sign of the gravitational
constant and also whether $K$ and $\Lambda $ are switched off or on, with $%
K=-1$, $m=1$ and $\protect\lambda =1$. A bounce is found, but in this case
there is no turnaround at a maximum and subsequent recollapse. }
\label{veff4}
\end{figure}

The most interesting follows from querying the sign of the gravitational
constant and also whether $K$ and $\Lambda $ are switched off or on (and set
to whatever non-zero value). Then, the Friedmann equation is: 
\begin{equation}
H^{2}=-\frac{K}{2a^{2}}+\frac{\lambda }{2}+S\sqrt{\frac{m^{2}}{a^{8}}+\frac{1%
}{4a^{4}}+\frac{\lambda ^{2}}{4}},
\end{equation}%
for the RHS of which we find a single zero at: 
\begin{equation}
a=\left( \frac{2m^{2}}{-K\lambda }\right) ^{\frac{1}{6}},
\end{equation}%
if $S=-1$ and $K\Lambda <0$. By drawing the effective potential (see Fig.~%
\ref{veff4}) we see that it corresponds to a bounce for $K=-1$, $\Lambda >0$
and $S=-1$. A bounce is found; however, in this case there is no turnaround
and recollapse. Nonetheless, the cosmological singularity appears to be
removed here -- no doubt due to the uncertainty in the sign of $G$ early on.
Note that, for $K=1$, $\Lambda <0$ and $S=1,$ there is also a root, but it
corresponds to a turnaround, since the effective potential is minus that in
Figure~\ref{veff4}. This would be no different from the standard case. 

\subsubsection{Case II}

\label{caseII}

It could also be that we cannot know simultaneously whether $K=\pm 1$, and
whether $G$ and $\Lambda $ are switched on or off. The sign issue therefore
concerns the spatial curvature $K$, which we know to be non-zero, but
quantum mechanically can be a superposition of $K=1$ and $K=-1$. Then, the
Friedmann equation resulting from the diagonalized hamiltonian becomes: 
\begin{equation}
H^{2}=\frac{m}{2a^{4}}+\frac{\lambda }{2}+S\sqrt{\frac{m^{2}}{4a^{8}}+\frac{1%
}{a^{4}}+\frac{\lambda ^{2}}{4}}.
\end{equation}%
Taking the $S=-1$ branch, we see that there are no solutions (the RHS is
negative) unless $m\lambda =2$ in which case 
\begin{equation}
H\equiv 0.
\end{equation}%
Therefore, we obtain static solutions for \textit{all} $a$, with indifferent
stability.

\subsubsection{Case III}

\begin{figure}[h]
\begin{center}
\scalebox{0.5}{\includegraphics{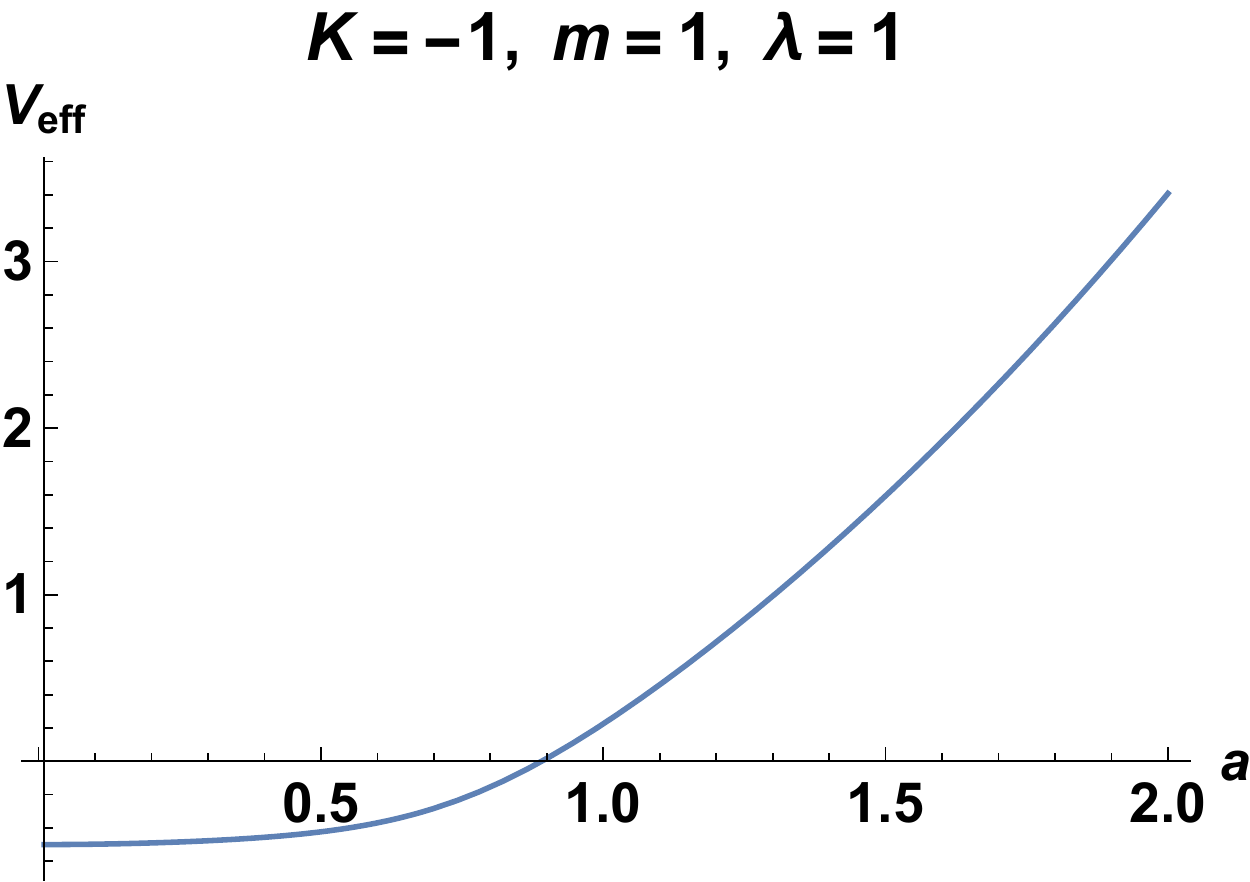}}
\end{center}
\caption{The effective potential when we query the sign of $\Lambda $ and
also whether or not $G$ and $K$ are switched on, with $K=-1$, $m=1$ and $%
\protect\lambda =1$. }
\label{veff5}
\end{figure}

Finally, let us query the sign of $\Lambda ,$ and whether or not $G$ and $K$
are switched on (with $G>0$ and either $K=1$ or $K=-1$). The Friedmann
equation becomes: 
\begin{equation}
H^{2}=\frac{m}{2a^{4}}-\frac{K}{2a^{2}}+S\sqrt{\frac{m^{2}}{4a^{8}}-\frac{1}{%
4a^{4}}+\lambda ^{2}}.
\end{equation}%
For the $S=-1$ branch, we have a single zero when $K=-1$: 
\begin{equation}
a=\left( \frac{-mK}{2\lambda ^{2}}\right) ^{\frac{1}{6}}.
\end{equation}%
However, no novelties are found in this case. Plotting the effective
potential (see Fig~\ref{veff5}) we see that this zero corresponds to a
turnaround. The singularity is still present. 

In summary, there are striking qualitative novelties in the dynamics if we
take the $S=-1$ state and use our formalism to switch on/off at least two of 
$K$, $G$ and $\Lambda$. One can see that in all other cases no new zeros for 
$\dot a$ are generated, the effects of the new dynamics limiting themselves
to subtle differences in the transition regions between the eras when one of
the 3 species dominates. If we take the $S=-1$ state, however, should we
probe the on/off switch of $K$, $G$ and $\Lambda$ we find generically an
early bounce. Ditto if we query the on/off switch of $K$ and $\Lambda $ and
the sign of $G$. The only other case of any novelty was discussed in Section~%
\ref{caseII} and is highly contrived.

\section{Discussion}

\label{disc} In this Section we briefly discuss the physical meaning of our
model within the broader context of theories of the constants of Nature. We
should clearly distinguish between \textit{variability} (e.g. changes in space
and time, or with energy, within a single universe) 
and \textit{diversity} (where the constants remain fixed in each
realization of the universe).

If we envisage theories with dynamical spacetime variations (for example
Refs.~\cite{BD, B, SBM}), then the variability (rather than the
instantaneous, local numerical values) of dimensionful constants is meaningful
because, even though the variations are tied to a system of units, its
choice is pegged down by the dynamics (e.g. the lagrangian) given to the
fields controlling the variations, typically rendered simple by that choice
of units. Seen in another way, the dynamics yields dimensionful integration
constants giving meaning to the variations. For example, in Brans-Dicke (BD)
theory \cite{BD}, the BD scalar field $\phi =G^{-1}$ is dimensionful but
satisfies a second-order differential equation whose two constants of
integration have the same dimensions, and their ratios with $\phi $ are
trivially dimensionless, and so there is no confusion over Brans-Dicke being
a theory for a varying dimensional constant ($G$).

Quite another situation arises when we imagine an ensemble of possible
universes, each with its own set of fixed values for the constants, which
then vary throughout the ensemble (such as in the Hilbert space proposed
here, if all the operators associated with the constants commute). Then,
their different numerical values are devoid of physical meaning if they
produce the same dimensionless constants, and amount to nothing more than a
rigid (i.e. fixed throughout each universe) change of units.

However, as this paper shows, we now see that, even in the latter context,
if dimensionful constants are associated with operators that do not commute,
the story changes. A physical meaning can be assigned to \textquotedblleft
diverse" dimensionful constants, because their complementarity precludes the
naive hamiltonian evolution associated with them. A different operational
dynamics arises from confronting the diversity of non-commuting constants.

A further contribution to this discussion can be extracted from the models
presented in Section~\ref{su2}. A change of units can never undo the
information about the sign of a constant, or whether it has a zero or a
non-zero value. This is true even before non-commutativity is applied to
these models.

\section{Conclusions}

\label{concs}

We have introduced a new approach to handling traditional physical constants
in a quantum cosmological setting (quite different from either of the traditional ones~\cite%
{quantum-cosmo,loop-cosmo1,loop-cosmo2,vsl-cosmo}) by elevating constants of
Nature to quantum operators in a kinematical Hilbert space. They do not have
classical counterparts, and their commutators do not arise from classical
Poisson brackets. Accordingly, the commutators of these constants do not
endow them with any time evolution: in this sense they are the genuine
constants -- although their observed values are not fixed to take only a
single set of values. We began by discussing the situation where the
operators representing constants were commuting. In this case our proposal
resembles the treatment given to classical parameters in quantum information
theory -- at least, if we live in an eigenstate of the constants. The
possibility of non-commutativity introduces novelties, due to its associated
uncertainty and complementarity principles -- and may even preclude
hamiltonian evolution. In this case the system typically evolves as a
quantum superposition of hamiltonian evolutions resulting from a
diagonalization process, and these are usually quite distinct from the
original one (defined in terms of the non-commuting constants).

We presented several toy examples targeting $G$, $c$ and $\Lambda $, in the
context of the dynamics of homogeneous and isotropic universes. If we base
our construction on the Heisenberg algebra and the quantum harmonic
oscillator, the alternative dynamics tend to silence matter (effectively
setting $G$ to zero), and make the spatial curvature and the cosmological
constant terms in the Friedmann equations act as if their signs are
reversed. As a result, the early universe expands as a quantum superposition
of different Milne or de Sitter metrics regardless of the material equation
of state, even though the matter nominally dominates the density (in $\rho $
but not $G\rho $). A superposition of Einstein static universes was obtained
as another worked example. We also investigated the results of basing our
construction on the algebra of $SU(2)$, into which we can insert information
about the sign of a constant of Nature, and whether its action is switched
on or off. In this case we found examples displaying quantum superpositions
of bounces in the initial state for the universe.

Our discussion has targeted the most interesting cases of quantising the
constants $c,G$ and $\Lambda $ because they have immediate controlling
consequences for the dynamics of homogeneous and isotropic universes. Other
investigations can be made of cosmological models with different dynamics
following the same quantum states of constants, where we have found that the
quantum behaviour of simple anisotropies with isotropic 3-curvature does not
reproduce that of massless scalar fields, as in the classical theory. In
order to extend this approach to explore other constants, like the fine
structure constant~\cite{B,SBM}, the electron-proton mass ratio~\cite{BMem}, or the parameters
defining the minimal standard model~\cite{KMSM,WSM}, choices have to be made about possible
non-commuting operator pairs. These extensions will be subjects for further
work.

Another possible extension of this work concerns the classical limit of
these theories, and what the wavefunction of the constants might be. It was
suggested in~\cite{notime} that the parameter $\mathfrak{h}$ controlling non-commutativity
(which need not be directly related to the usual $\hbar$) could be 
a function of the cosmic density ($\mathfrak{h}=\mathfrak{h}(\rho )$) and that
below a certain density it could be pushed to zero, $\mathfrak{h}\rightarrow 0$, possibly at a
phase transition. This could assist in bringing about classicality, but it
might also require the wavefunction to be a coherent state in  the original $\kappa _{1}$
and $\kappa _{2}$ in (\ref{coms3}). 
The detailed construction of these states, however, is not straightforward. Note that
they are {\it not} the naive coherent states built from $\alpha $ defined from (\ref{alpha1}) and (\ref{alpha2}), in fact,
the latter are time dependent and form squeezed states in $\kappa _{1}$
and $\kappa _{2}$ at late times. This is because the transformation between the $\alpha_i$ and the $\kappa_i$
is time-dependent, with implications to be more thoroughly studied in a paper in preparation.

If the late-time dynamics is brought about by a phase transition
which sets $\plk=0$, then no prediction is made for the late-time value of the constants (it is simply put in by hand
when the appropriate coherent state is built). Likewise, there is {\it no} implication that a fundamental indeterminacy between constants might be at work at late times. However, the phase transition scenario is simply the minimal model, and further complexity could be built in. The issue of bringing about the late-time dynamics in our model would then be similar to the graceful exit problem in inflation, or the equivalent for all alternative scenarios. In such non-minimal models it is possible that the formalism proposed here would be at work nowadays, with interesting phenomenological implications. Such matters are left for further work. Note also that the issue of the collapse of the wave-function does not need to be addressed unless we are ready to embraced such non-minimal models.

To conclude we should stress that our constructions are not currently supported by any theory aspiring to the status
of ``fundamental'' (such as string theory or loop quantum gravity).  For example, there is no known fundamental principle specifying which constants should be included in the algebra of $SU(N)$  
generalizing Section~\ref{su2}. We are taking the first steps in a new direction and many questions are raised that we cannot yet answer. We hope that others will add to our beginnings and fill these gaps. In particular it would be interesting to see whether quantum gravity theories have anything to say in this respect. Although the discussion of matter coupling constants 
is hampered by the fact that most formalisms apply to vacuum dynamics only, a discussion of 
$\Lambda$ and $G$ should be possible. Some hints in this direction can be found in~\cite{quantumlambda,LeeGc}.

\section{Acknowledgements}

We thank Stephon Alexander, David Jennings and Lee Smolin for discussions on matters related to
this paper and Remo Garattini and Sabine Hossenfelder for bringing to our attention the connection 
between their work and ours. 
JDB and JM were both supported by consolidated grants from the
Science and Technology Facilities Council (STFC) of the UK.



\begin{thebibliography}{99}
\bibitem{ein} A. Einstein, Letter to I. Rosenthal-Schneider (30 March 1947),
English translation in I. Rosenthal-Schneider, \textit{Reality and
Scientific Truth}, Wayne State Press, Detroit, 1980, pp. 56-7 and J.D.
Barrow, \textit{The Constants of Nature, }Bodley Head, London (2002), pp.
33-42.

\bibitem{swamp} H. Ooguri and C. Vafa, Nucl. Phys. B \textbf{766,} 21
(2007); C. Vafa, The string landscape and the swampland,
arXiv:hep-th/0509212; T. D. Brennan, F. Carta, and C. Vafa, The String
Landscape, the Swampland, and the Missing Corner, arXiv:1711.00864

\bibitem{jdb2} J.D. Barrow, Phys. Rev. D \textbf{71}, 083520 (2005)

\bibitem{3d} P. G. O. Freund, Nucl. Phys. B209, 146 (1982); E. W. Kolb, M.
J. Perry, and T. P. Walker, Phys. Rev. D \textbf{33}, 869 (1986); J.D.
Barrow, Phys. Rev. D \textbf{35}, 1805 (1987)

\bibitem{webb} J.K. Webb, M.T. Murphy, V.V. Flambaum, V.A. Dzuba, J.D.
Barrow, C.W. Churchill, J.X. Prochaska, and A.M. Wolfe. Phys. Rev. Lett. 
\textbf{87}, 091301 (2001); J.A. King, J.K. Webb, M.T. Murphy, V.V.
Flambaum, R.F. Carswell, M.B. Bainbridge, M.R. Wilczynska, and F.E. Koch,
Mon. Not. Roy. astron. Soc. \textbf{422}, 3370 (2012)

\bibitem{BD} C. Brans and R.H. Dicke, Phys. Rev. \textbf{124}, 925 (1961)

\bibitem{B} J.D. Bekenstein, Phys. Rev. D \textbf{25}, 1527 (1982)

\bibitem{SBM} H.B. Sandvik, J.D. Barrow and J. Magueijo, Phys. Rev. Lett. 
\textbf{88}, 031302 (2002)

\bibitem{Qinf} G. Chiribella and R.W. Spekkens, (eds), \textit{Quantum
Theory: Informational Foundations and Foils}, Springer, New York (2016)

\bibitem{quantumlambda} S.~Alexander, J.~Magueijo and L.~Smolin, The quantum
cosmological constant, arXiv:1807.01381

\bibitem{notime} J.~Magueijo and L.~Smolin, A universe that does not know
the time, arXiv:1807.01520

\bibitem{remo1}
  R.~Garattini,
  J.\ Phys.\ A {\bf 39}, 6393 (2006)
  doi:10.1088/0305-4470/39/21/S33
  [gr-qc/0510061].

\bibitem{multi} B. Carr, (ed.) Universe or Multiverse?, Cambridge UP, Cambridge, (2007)

\bibitem{sabine}
  S.~Hossenfelder,
  Found.\ Phys.\  {\bf 42}, 1452 (2012)
  doi:10.1007/s10701-012-9678-0
  [arXiv:1207.1002 [gr-qc]].

\bibitem{JDBOb} J.D. Barrow, Observatory \textbf{113}, 210 (1993)

\bibitem{static} J.D. Barrow, G.F.R. Ellis, R. Maartens, and C.G. Tsagas,
Class. Quantum Gravity \textbf{20}, L155 (2003).

\bibitem{quantum-cosmo}  J.~J.~Halliwell, 
In \textit{Jerusalem 1989, Proceedings, Quantum cosmology and baby universes}
159-243 and MIT Cambridge - CTP-1845, gr-qc/0909.2566 

\bibitem{loop-cosmo1}  A.~Ashtekar, Lect.\ Notes Phys.\ \textbf{863}, 31
(2013)  

\bibitem{loop-cosmo2}  M.~Bojowald,  Rept.\ Prog.\ Phys.\ \textbf{78},
023901 (2015)

\bibitem{vsl-cosmo}  A.~Balcerzak,  JCAP \textbf{1504}, 019 (2015). 

\bibitem{BMem} 
  J.~D.~Barrow and J.~Magueijo,
  Phys.\ Rev.\ D {\bf 72}, 043521 (2005)
  doi:10.1103/PhysRevD.72.043521
  [astro-ph/0503222].

\bibitem{KMSM} 
  D.~Kimberly and J.~Magueijo,
  Phys.\ Lett.\ B {\bf 584}, 8 (2004)
  doi:10.1016/j.physletb.2004.01.050
  [hep-ph/0310030].

\bibitem{WSM} 
  F.~Wilczek,
  arXiv:0708.4361 [hep-ph].

\bibitem{LeeGc} 
  L.~Smolin,
  Class.\ Quant.\ Grav.\  {\bf 33}, no. 2, 025011 (2016), 
  [arXiv:1507.01229 [hep-th]].




\end{thebibliography}
\end{document}